# Traffic Modeling and Forecast based on Stochastic Cell-Automata and Distributed Fiber-Optic Sensing – A Numerical Experiment

Yoshiyuki Yajima and Takahiro Kumura

**Abstract**— This paper demonstrates accurate traffic modeling and forecast using stochastic cell-automata (CA) and distributed fiber-optic sensing (DFOS). Traffic congestion is a dominant issue for managers and companies of highways. To reduce congestion, real-time traffic control by short-term forecast is necessary. For achieving this, data assimilation using a stochastic CA model and DFOS is promising. Data assimilation with a CA enables us to model real-time traffic flow with simple processes even when rare or sudden events occur, which is challenging for usual machine learning-based methods. DFOS overcomes issues of conventional point sensors that have dead zones of observation. By estimating optimal model parameters that reproduce observed traffic flow in the simulation, future traffic flow is forecasted from the simulation. We propose an optimal model parameter estimation method using mean velocity as an extracted feature and the particle filter. In addition, an estimation methodology for the microscopic traffic situation is developed to set the initial condition of simulation for forecast in accordance with observation. The proposed methods are verified by simulation-based traffic flow with arbitrarily set model parameters as emulated DFOS data. The simulation uses the stochastic Nishinari-Fukui-Schadschneider model, a stochastic cell-automata traffic model. The optimal model parameters are successfully derived from posterior probability distributions (PPDs) estimated from the proposed method and emulated DFOS data. In contrast, optimal model parameters estimated emulated point sensor data fails to identify the grand-truth values. The estimated PPDs of model parameters also indicate that each parameter has different sensitivities to observed mean velocity. A short-term traffic forecast up to 60 minutes in the future is also carried out. Using optimal model parameters estimated from DFOS, the forecast error of 1 km sections and 1-minute mean velocity is approximately $\pm 10$ km/h (percentage error is 18%). These errors attain half of those when conventional point sensors are used. We conclude that DFOS is a powerful technique for traffic modeling and short-term forecast because of its continuous data along the road without dead zones.

## 1 INTRODUCTION

Traffic congestion and jams are severe issues in many countries. The economic loss of congestion due to wasting time and fuel is non-negligible. For instance, it was as huge as 87 billion USD in 2018 in the USA [1]. In addition, risks of traffic accidents significantly increase when congestions occur due to high vehicle density. Drives feel stress when they experience congestions, which decrease service quality by highway managers and companies. In the era of autonomous vehicles, vehicle-infrastructure integration will be standard. New technologies are essential to acquire traffic flow data from roads, reduce congestion from appropriate control, and achieve free flow as much as possible.

It is necessary to recognize signs of congestion as soon as possible and control traffic flow from an appropriate traffic forecast in order to moderate harmful effects. To achieve this goal, an accurate traffic forecast gives us a solution. Traffic forecast is recently machine-learning-based method is commonly used (e.g., [2]). These methods learn patterns and trends of historical data and forecast future traffic flow according to data similar to the current situation. However, such approach has a weak point. Since they are based on the most effective feature in the extensive historical data, forecast accuracy decreases when rare and sudden events occur. For example, the probability that traffic accidents occur is very low. Machine-learning-based traffic forecast models are challenging to learn such unusual cases because they learn the ordinaries. Hence, the forecast can be largely deviated from the actual traffic flow, particularly shortly after congestions occur where congestions are still small. A forecast approach that considers the recent and real-time traffic situation is promising.

*Yoshiyuki Yajima, Takahiro Kumura, Visual Intelligence Research Laboratories, NEC Corporation in Shimonumabe 1753, Nakahara-ku, Kawasaki, 211-8666 Kanagawa, Japan. Email: yoshiyuki-yajima@nec.com*

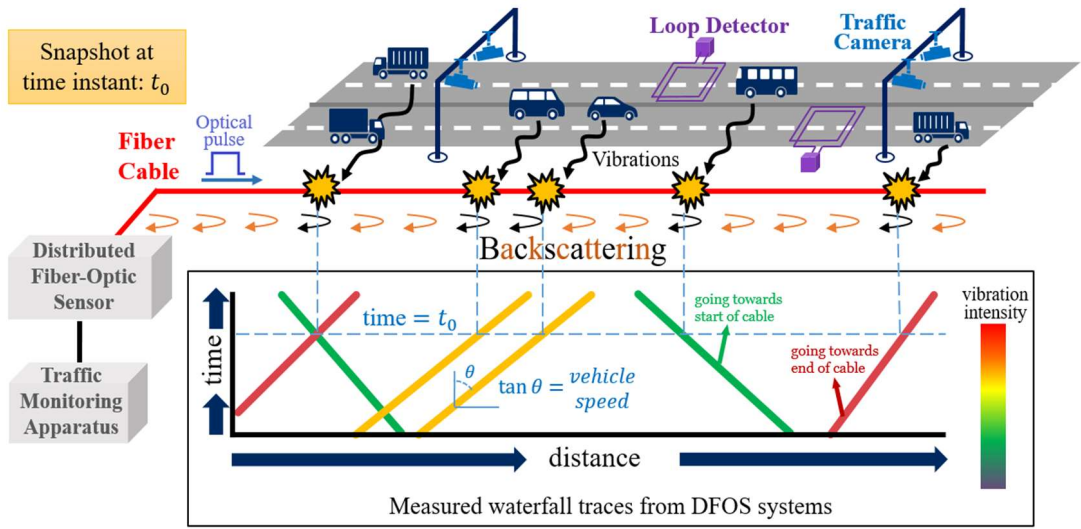

Figure 1. Schematic view of traffic monitoring using DFOS. It detects vibrations along optical fibers. Therefore, vehicle trajectories are visualized in the time-space plane by tracking the vibration of driving vehicles without dead zones. In contrast, the observable area of traditional point sensors like cameras and loop detectors is limited. Slopes of trajectories correspond to vehicle velocities. For instance, a trajectory of a high-velocity vehicle is shallow, and a low-velocity vehicle shows a steep trajectory. Using these trajectories, mean velocities in a road section are obtained.

To overcome the issue, we propose data assimilation approach using a theoretical traffic model. Data assimilation calibrates theoretical model parameters to match the simulation by the model and recent observed results. The simulation of the calibrated model parameters forecasts future traffic flow. Specifically, the parameters of the theoretical model are optimized to reproduce observed data. Then, future traffic flow is simulated using the optimal model parameters.

This paper adopts a stochastic cell-automata (CA) as a theoretical traffic model. Traffic models based on CA are Lagrangian-specification fluid dynamics and each vehicle's behavior is simulated. They are also a self-organization system; simple rules in CA generate complex phenomena and realistic traffic flow observed in the real world. In addition to simple processes that lead to easiness of implementation in the simulation program, improved computational machine power promotes the use of CA in the traffic modeling [3], as well as other theoretical models based on the density-wave theory (e.g., the Aw-Rascle-Zhang model [4, 5]) and car-following (or optimal velocity) model (e.g., [6]).

A primitive CA-based traffic model is proposed by Nagel and Schreckenberg [7], known as the NS model. They introduced a probabilistic brake to reproduce the instability of traffic flow. Nishinari et al. [8] suggested the traffic model introducing the quick start [9, 10] and slow-to-start effect [11, 12]. The former considers drivers' anticipation based on the vehicle driving two cars ahead. The latter considers the recognition delay of drivers and the inertia of vehicles. It is known as the Nishinari-Fukui-Schadschneider (NFS) model. Sakai et al. [13] expands the NFS model by adding the random brake effect and the stochastic quick-start and slow-to-start effect, i.e., the stochastic-NFS (S-NFS) model.

There is another issue for modeling traffic flow with conventional monitoring equipment, such as video cameras and traffic counters. These point sensors can observe traffic flow only at positions where they are located. In other words, they cannot observe congestion and traffic accidents occurring between sensors. In Japan, their spacing is 200-500 m in urban expressways and 2 km in interurban expressways. This issue limits observed data and makes achieving accurate data assimilation challenging.

To overcome the issue of point sensors, we propose to use distributed fiber-optic sensing (DFOS). DFOS is a real-time sensing technology that monitors vibration along the optical fiber using back-scattering light. The fiber itself becomes a one-dimensional continuous sensor. When optical fibers are already installed along highways for telecommunications, vibrations from driving vehicles are detected and the paths of the vehicles are visualized, as shown in Figure 1. Each diagonal path in the time-space plane is the trajectory of each vehicle, and its slope corresponds to velocity.

Traffic monitoring using DFOS was first tired by Wellbrock et al. [14]. Narisetty et al. [15, 16] developed a deep neural network that recognizes trajectories and derives space mean velocity (hereafter, mean velocity) of a road section (e.g., 1 km) and a unit time (e.g., 1 minute). This system is already used for real-time congestion monitoring in some expressways in Japan. Additional denoising and localization of fibers to actual road location methods also have been developing [17, 18, 19, 20, 21].

This paper validates that DFOS enables more accurate model parameter estimation and traffic forecast than conventional point sensors using synthetic data. Section 2 proposes our methodology to estimate model parameters and traffic state from observed data for setting the initial condition for the forecast. Section 3 describes simulation settings to verify our methods. Results and their discussion of the model parameter estimation and traffic forecast are presented in section 4. Section 5 concludes this paper.

## 2 METHOD

### 2.1 Model Parameter Estimation

Our proposed method to estimate model parameters is based on the particle filter. The particle filter is a generalized state-space model widely used in time-series analysis. In this case, observed variables are mean velocity and state variables are model parameters. The reason not trajectories but mean velocity is used is a feature extraction of DFOS data. By adopting the particle filter, posterior probability distributions (PPDs) of each model parameter are estimated. The detailed process is as follows.

At the initial state $t = 0$, $N$ sets of model parameters are sampled from prior probability densities of each parameter. When we do not have prior information, each parameter set is sampled from the parameter space by grid search. At each sampled point in the parameter space, particles are put. We denote particles at $n$-th parameter set as $\{x_{n,t=0}^{(i)}\}$.

At time $t$ and $n$-th parameter set $\boldsymbol{\theta}_n$, error function $E_{n,m,t}(\boldsymbol{\theta}_n)$ is defined as the percentage error of simulated mean velocity,

$$E_{n,m,t}(\boldsymbol{\theta}_n) \equiv \frac{100\left|v_{n,m,t}^{\text{sim}}(\boldsymbol{\theta}_n) - v_{m,t}^{\text{obs}}\right|}{v_{m,t}^{\text{obs}}}, \tag{1}$$

where $v_{n,m,t}^{\text{sim}}$ is the simulated mean velocity at $m$-th road section in $n$-th parameter set and $v_{m,t}^{\text{obs}}$ is observed mean velocity at the same road section and the same time. The percentage error makes the penalty severe for observed low-velocity sections. For example, percentage error of 10% and $v^{\text{obs}} = 100$ km/h allows the error of $v^{\text{sim}} = \pm 10$ km/h whereas $v^{\text{obs}} = 30$ km/h allows the simulation error of only $\pm 3$ km/h. The error is low if the simulation result by $\boldsymbol{\theta}_n$ successfully reproduces observed congestion.

The likelihood $\mathcal{L}_{n,t}(\boldsymbol{\theta}_n)$ is defined as

$$\mathcal{L}_{n,t}(\boldsymbol{\theta}_n) \equiv \prod_m \frac{1}{\sqrt{2\pi\sigma^2}} \exp\left[-\frac{E_{n,m,t}^2(\boldsymbol{\theta}_n)}{2\sigma^2}\right], \tag{2}$$

where $\sigma$ is the hyperparameter that controls the tolerance of discrepancy between observed and simulated mean velocity. The product of the normal distributions for the road sections $m$ also makes the penalty severe if locations of congestion in the simulation are inconsistent with observed results. Using the likelihood, weight of $\boldsymbol{\theta}_n$ is defined as

$$w_{n,t}(\boldsymbol{\theta}_n) \equiv \left[\ln \mathcal{L}_{n,t}(\boldsymbol{\theta}_n)\right]^{-2}. \tag{3}$$

Note that since $\mathcal{L}_{n,t}(\boldsymbol{\theta}_n)$ is a product of the values of the probability density function, $\ln \mathcal{L}_{n,t}(\boldsymbol{\theta}_n)$ is always negative. Therefore, converting the inverse square of the logarithmic likelihood always returns positive and becomes higher when the observed and simulated mean velocities by $\boldsymbol{\theta}_n$ are similar. After $w_{n,t}(\boldsymbol{\theta}_n)$ for all $N$ parameter sets is obtained, each weight is normalized,

$$w_{n,t}(\boldsymbol{\theta}_n) \leftarrow \frac{w_{n,t}(\boldsymbol{\theta}_n)}{\sum_{n=1}^{N} w_{n,t}(\boldsymbol{\theta}_n)}. \tag{4}$$

Using $w_{n,t}(\boldsymbol{\theta}_n)$, new particles $\{x_{n,t}^{(i)}\}$ are re-sampled with the statistical weight of $w_{n,t}(\boldsymbol{\theta}_n)$. As a result, the number of particles at $\boldsymbol{\theta}_n$ that reproduce observed mean velocities increases. Then moving to the next time $t+1$ and the same process is applied. Posterior probability distributions of each parameter are obtained from histograms of $w_{n,t}(\boldsymbol{\theta}_n)$ in the final time step. Optimized model parameters are identified from the maximum a posteriori (MAP) or the expectation of the posterior probability distributions. As shown in section 4, PPDs of model parameters derived from this method converge well using tens of a few minutes of observed mean velocity data. Therefore, the proposed method does not need long-term observed data in contrast to ordinal machine-learning-based methods recognizing trends and patterns.

## 2.2 Traffic Forecast by Simulation and Setting of the Initial Condition

Future traffic flow is forecasted by the simulation, which adopts optimal model parameters estimated by the proposed method in section 2.1. This forecast assumes that the model parameters are the same as those in the observed traffic flow. Therefore, the forecast is valid as long as the drivers' behavior and the road situation are unchanged. It indicates that this methodology is a short-term forecast from tens of minutes up to an hour rather than a long-term forecast on a time scale of several days and months.

Initial conditions have to be appropriately set considering the latest observed traffic situation for forecasting. To do this, the initial condition is estimated based on the latest observed mean velocities to set it, considering the latest situation appropriately. As the first step, the vehicle number in each 1 km section is estimated from mean velocity according to the fundamental diagram of density–velocity, i.e., $k$–$v$ relation. The relation is obtained from the simulation result and a theoretical model at the optimal model parameters. This paper adopts the Underwood model [22] as the theoretical model. Based on the relation fitted with the Underwood model, mean density, the total number of vehicles in each 1 km section is estimated from the latest observed mean velocity. The road section, including the bottleneck, should differ from other sections regarding fitting results. Therefore, the fitting is performed depending on the existence of the bottleneck. After the total number of vehicles is obtained, each car position is determined by assuming the existence probability in the fast lane and the distance between each vehicle is equal in each lane.

Once the positions of each vehicle are determined, the velocity of each vehicle is estimated from the observed mean velocity and the estimated total number of vehicles. Here, we assume velocity dispersion is minimal because the probability of traffic accidents is very low. For instance, when the observed mean velocity is 70.2 km/h in a 1 km section, it is assumed that vehicles driving at only 60 and 80 km/h exist in the section when the velocity resolution of the CA setting is 20 km/h. By considering that the harmonic mean of each vehicle velocity is equal to the observed mean velocity, the number of vehicles driving at the lower velocity (in the case above, it is 60 km/h) $N_l$ can be derived from the following equation,

$$N_l = \lfloor 0.5 + Nv_l(v_l + dv - v_s)/v_s dv \rfloor, \tag{5}$$

where $N$ is the total number of vehicles in the section, $v_l$ is the lower velocity, $dv$ is the velocity resolution, $v_s$ is the mean velocity, and $\lfloor . \rfloor$ is the floor function. Then, each velocity and position of the vehicle are randomly combined. Processes so far are applied for all road sections where mean velocity is measured. In this way, the initial condition considering the latest observational results is generated.

This method for the initial condition setting enables estimating the latest microscopic traffic condition in the simulation without any trial-and-error. This process is necessary for the traffic simulation using CA models because they are microscopic traffic models. In a congestion case, many slow vehicles, in accordance with the observed mean velocity, are placed in the simulation model in the road section where congestion is observed. Even in a rare case of sudden congestion by an accident, it can be reflected in the simulation model, which is impossible in the machine-learning-based method. An accurate traffic forecast is attained because of the appropriate initial condition of the simulation.

## 3 NUMERICAL EXPERIMENTS

### 3.1 Simulation for Model Parameter Estimation

Our proposed methods for model parameter estimation and traffic forecast are verified using simulated mean velocities as pseudo-measurement data of DFOS. The traffic simulation is based on the S-NFS model. Updating processes of vehicle velocity and position of the S-NFS model are as follows;

Step 1. Acceleration

$$v_i^{(1)} = \min\{V_{\max}, v_i^t + 1\}, \tag{6}$$

Step 2. Slow-to-start effect

$$v_i^{(2)} = \min\{v_i^{(1)}, x_{i+S}^{t-1} - x_i^{t-1} - S\} \text{ with probability } q \tag{7}$$

Step 3. Quick start effect

$$v_i^{(3)} = \min\{v_i^{(2)}, x_{i+S}^t - x_i^t - S\} \tag{8}$$

Step 4. Random brake effect

$$v_i^{(4)} = \max\{0, v_i^{(3)} - 1\} \text{ with probability } p \tag{9}$$

Step 5. Avoidance of collision

$$v_i^{t+1} = \min\{v_i^{(4)}, x_{i+1}^t - x_i^t - 1 + v_{i+1}^{(4)}\} \tag{10}$$

Step 6. Moving

$$x_i^{t+1} = x_i^t + v_i^{t+1}, \tag{11}$$

where $v_i^t$ is velocity of $i$-th vehicle from the origin of the road at time $t$ and $x_i^t$ is its position. The number of a forward vehicle the driver consider $S$ is defined as

$$S = \begin{cases} 1 \text{ with probability } 1 - r, \\ 2 \text{ with probability } r. \end{cases} \tag{12}$$

Namely, the driver considers the distance to the two-forward vehicle with the probability $r$. The probabilities $p$, $q$, and $r$ are features of traffic flow. If $p$ and $q$ are high, traffic flow is unstable (tends to be congested). If $r$ is high, the traffic capacity of the road is high, although the critical density to transform free flow into congestion is low. Thus, traffic flow in low $r$ is stable whereas mean velocity decreases when flow rate is high. By optimizing these three parameters, observed traffic flow can be modeled in the S-NFS model.

The length of each cell and time-step size are 10 m and 1.8 seconds, respectively, which correspond to the velocity resolution of 20 km/h. The road model is 10 km in length and has two lanes. The speed limit in the fast and slow lane is 100 and 80 km/h, respectively. These are common in expressways of Japan. As the bottleneck, the lower speed limit $V^{\text{BN}}$ is applied in the 8.4–8.6 km section of the road. $V^{\text{BN}}$ is set in three patterns: 20, 40, and 60 km/h. Therefore, the number of components of the model parameter set $\boldsymbol{\theta}$ is 4; $\boldsymbol{\theta}_n = (V_n^{\text{BN}}, p_n, q_n, r_n)$. In the simulations, random brake probability $p$ is sampled from the range of 0.05–0.6 in 0.05 increments. Slow-to-start probability $q$ and anticipation probability $r$ are sampled from 0.1–0.8 in 0.1 increments and 0.75–0.99 in 0.03 increments, respectively. Therefore, the number of model parameter sets $N$ in Equation (4) is 2592.

The lane change algorithm is as follows. In each time step, vehicles check the maximum velocity that can be achieved considering the distance to forward cars and the speed limit in the current and adjacent lanes. Suppose the maximum velocity in the adjacent lane is higher than that in the current lane. In that case, the vehicle checks the velocity of the nearest rear vehicle in the adjacent lane and the distance to it. If the velocity difference and distance to the nearest rear vehicle are sufficient to avoid collision, the vehicle moves to the adjacent lane with a probability. The probability of the lane change is set as 10%. In this setting, the probability that a vehicle changes the lane at least once within 33 time steps (~1 minute) is 96.9% when the conditions of the lane change are always satisfied during that time.

The inflow rate at the origin of the road is set based on the typical value of observations in major 2-lane expressways in Japan. The time series inflow rate in the simulations is shown in Figure 2. This data is generated from the linear trend and random noise whose standard deviation is the same as observations in actual expressways. The linear trend is derived from an observation result for 3 hours in the morning when the traffic volume gradually increases. The probability of entering vehicles being placed in the fast lane is set as 60% based on actual expressways. As the initial condition, vehicles are placed 200 m apart in each lane at 80 km/h velocity. Simulations are performed for 20 minutes (668 time steps), which correspond to $t = -20$–$0$ minutes in Figure 2.

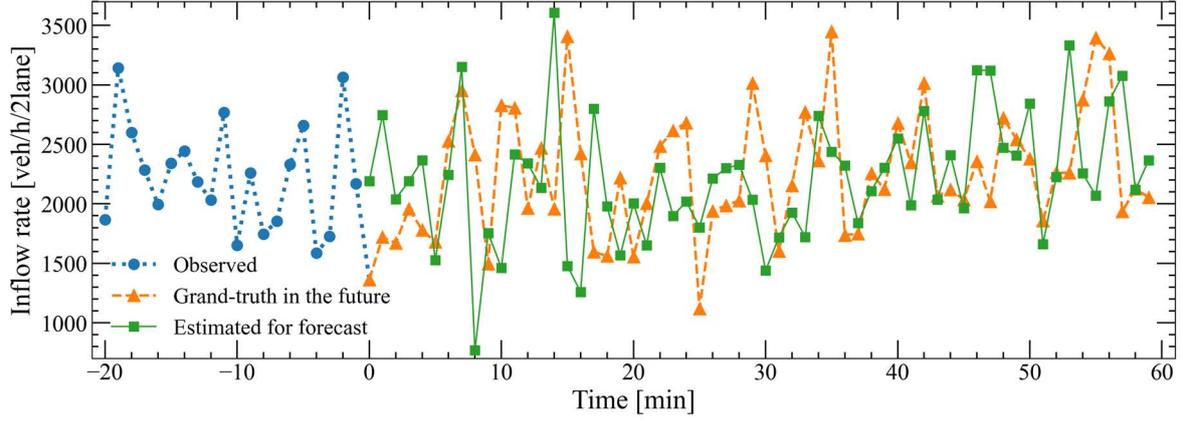

Figure 2. Time series data of inflow rate at the origin of the road. The present time is $t = 0$ minutes, i.e., the inflow rates of $t < 0$ minutes shown in dotted blue are already observed. The inflow rates in $t > 0$ are not yet known, therefore, the future inflow rate for the forecast indicated in solid green is estimated. Time series data in dashed orange is the grand-truth (actual) inflow data in the future.

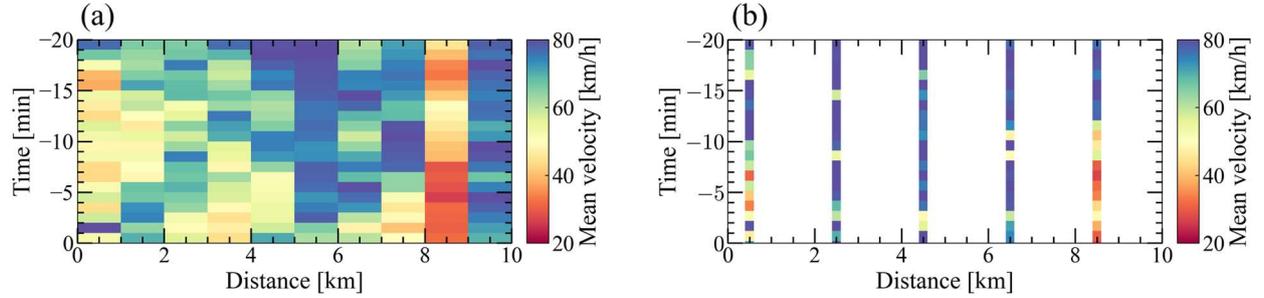

Figure 3. (a) Mean velocities with the given model parameters as pseudo-measured data using DFOS. (b) The same as (a) but for observed results by point sensors. Note that the observed areas are enhanced in the figure for easy understanding. However, the observable range of point sensors is approximately a few meters.

The simulation output is the trajectories of each vehicle on the time-space plane. From these trajectories, mean velocity averaged over 1 minute, each 1-km road section is derived based on the Edie method [23]. This output is the same as that in the traffic monitoring system using DFOS [15, 16]. As the grand-truth data, mean velocities with ($V^{BN}$, $p$, $q$, $r$) = (40 km/h, 0.36, 0.12, 0.98) are used as pseudo-measured data of DFOS. Figure 3 (a) shows the mean velocities of $t = -20$ to 0 minute in the grand-truth parameters. Our proposed method to estimate model parameters is verified using this pseudo-measured data.

To justify the usefulness of DFOS for data assimilation of traffic flow, results of model parameter estimation and forecast obtained from conventional point sensors are compared to those from DFOS. Therefore, pseudo-measured data of point sensors are also derived. In the road model of the simulation, point sensors are supposed to be located at 0.3, 2.3, 4.3, 6.3, and 8.3 km. The 2-km spacing is a typical situation in expressways in Japan. Mean velocity averaged over 1 minute is derived from the harmonic mean of instant vehicle velocities at sensor locations. Figure 3 (b) shows the mean velocity derived from point sensors.

### 3.2 Simulation for Forecast

The simulation for short-term forecast is carried out by adopting the optimal model parameters. The initial condition is set based on the proposed method in section 2.2 and the latest observed mean velocities at $t = -1$ to 0 minutes shown in Figure 3 (a). Since the 8–9 km road section includes the bottleneck, vehicle positions and velocities are estimated from the $k$–$v$ relation fitted by simulation data used only in the section. In other road sections, the fitted relation excluding simulation data of the 8–9 km section is used to set the initial condition. The existence probability is assumed to be 60% based on the typical value observed in Japanese expressways. Our proposed method cannot generate the initial condition of forecast simulation when using point sensors. Therefore, the final state at $t = 0$ minute at the optimal model parameters estimated from point sensor data is adopted as the initial condition for the forecast.

The inflow rate at the origin of the road in the future ($t > 0$ minute) has to be estimated. For simplicity, it is assumed that the linear trend adding random noise whose standard deviation is the same as that in observed data in actual Japanese expressways. It is shown in the green solid line of Figure 2. Based on the initial condition and estimated inflow rate time-series data, the traffic condition is simulated up to $t = 60$ minutes (2000 time steps).

## 4 RESULTS AND DISCUSSION

Figure 4 (a) shows PPDs of $V^{BN}$, $p$, $q$, and $r$ derived from the proposed method in section 2.1, their expectations, and the grand truth. High posterior probability indicates that parameters at the value well reproduce observed mean velocities shown in Figure 3 (a). Note that since $V^{BN}$ is discrete owing to velocity resolution, the expectation of PPD for $V^{BN}$ is not shown. Table 1 also summarizes MAP values, expectations, and grand-truth values. PPDs of the model parameters derived from point sensor data in the same way as section 2.1 are shown in Figure 4 (b) and their representative values are in Table 1. Although expectations by DFOS and point sensors are not so different, MAP values of DFOS are much closer to the grand-truth values than those of point sensors. Hence, MAP values are hereafter adopted as the optimal model parameters.

Compared with $V^{BN}$, $p$, and $r$, the accuracy of estimated $q$ is relatively worse. In addition, the PPD of $q$ does not show a clear peak unlike other PPDs. These results indicate that the sensitivity of $q$ to mean velocity is lower than those of $V^{BN}$, $p$, and $r$. In other words, the optimal value of $q$ is difficult to estimate, whereas observed traffic flow can be appropriately modeled using the proposed method. Optimal values of influential parameters $V^{BN}$, $p$, and $r$ can be identified. According to Sakai et al. 2006 [13], $q$ seems to only control the behavior of meta-stable states in the flow-density fundamental diagram. When $q$ is low (but non-zero), meta-stable states continue for a relatively long term. On the other hand, meta-stable states quickly turn into the congestion flow when $q$ is high. In contrast, $p$ and $r$ control mean vehicle speed through random braking and distance to the forward car ($r$ directly controls the traffic capacity of the road, while the velocities of vehicles decrease when inflow is higher than the capacity). $V^{BN}$ controls the scale

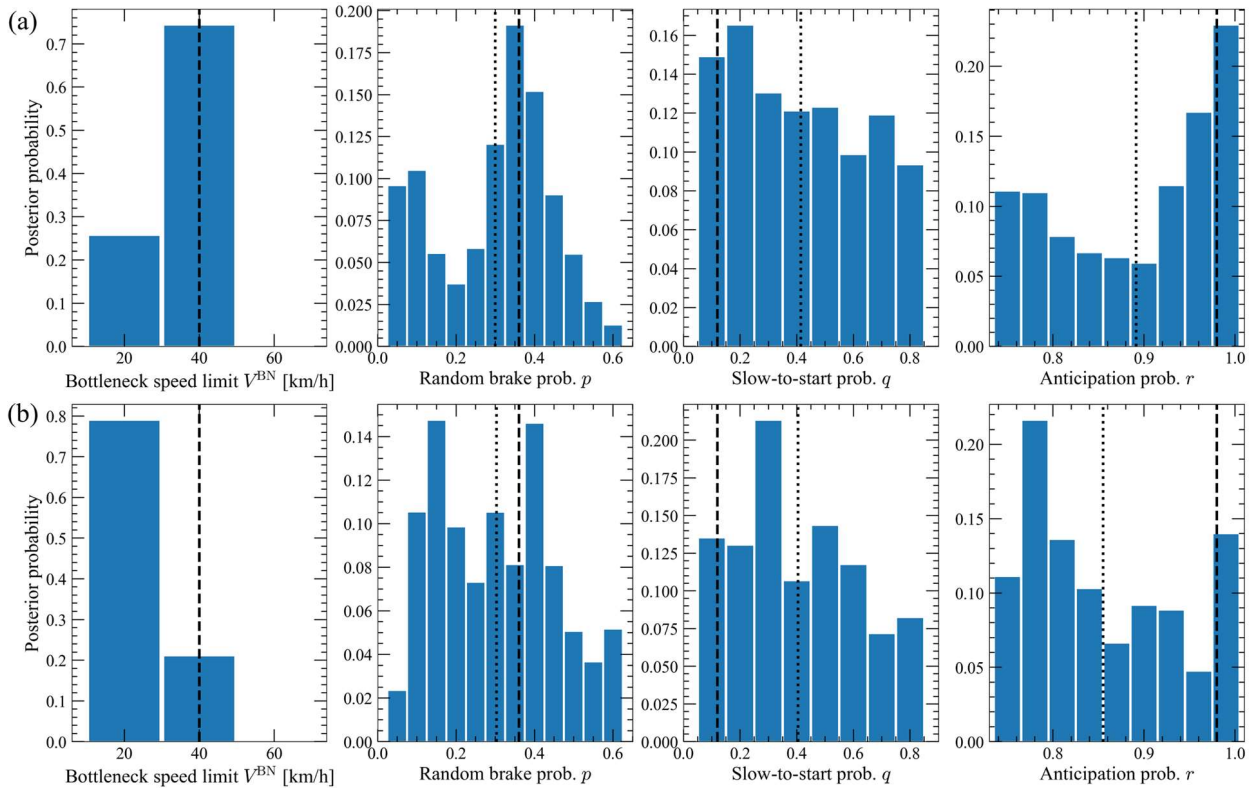

Figure 4. (a) PPDs of the model parameters derived from DFOS. (b) The same as (a) but for PPDs derived from point sensors. Vertical dashed and dotted lines indicate grand-truth values and expectations of PPDs.

TABLE 1.

MAP and expectations of model parameters estimated from DFOS, and grand-truth value.

|  | MAP (DFOS) | MAP (point sensor) | Expectation (DFOS) | Expectation (point sensor) | Grand truth |
| --- | --- | --- | --- | --- | --- |
| Bottleneck speed limit $V^{BN}$ | 40 km/h | 20 km/h | – | – | 40 km/h |
| Random brake prob. $p$ | 0.35 | 0.15 | 0.30 | 0.30 | 0.36 |
| Slow-to-start prob. $q$ | 0.20 | 0.30 | 0.41 | 0.40 | 0.12 |
| Anticipation prob. $r$ | 0.99 | 0.78 | 0.89 | 0.86 | 0.98 |

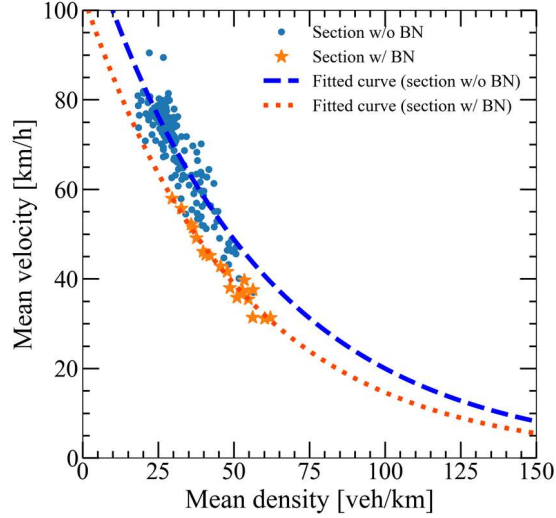

Figure 5. The density–velocity relation obtained from the simulation result of the MAP model parameters of DFOS data.

and growth speed of the traffic jam. Thus, $V^{BN}$, $p$, and $r$ exert a strong influence on mean velocity.

PPDs of $p$ and $r$ show bimodality. They are caused by the two possibilities of $V^{BN}$. The primary component of $p \sim$ 0.35 and $r \sim 0.99$ corresponds to $V^{BN} = 40$ km/h. The secondary component of $p \sim 0.1$ and $r \sim 0.75$ corresponds to $V^{BN} = 20$ km/h. When the weak bottleneck is accepted, model parameters for unstable traffic flow (high $p$ and $r$) are accepted. In contrast, the strong bottleneck increases the probability of model parameters in which traffic flow is stable (low $p$ and $r$).

Using these mean velocities from point sensors, PPDs of the model parameters are derived using the proposed method. Figure 4 (b) shows those PPDs. Unlike the results from DFOS, the peak values of the PPDs derived from point sensors are lower. It indicates that uncertainties become high due to limited information. Moreover, the MAP result of $V^{BN}$ is underestimated. This causes random brake probability $p$ and anticipation probability $r$ to be underestimated when these MAPs are adopted as optimal values. We conclude that the model parameters that reproduce observed traffic flow are successfully estimated using DFOS thanks to its continuous data, however, conventional point sensors cannot do so.

As a consequence of accurate parameter estimation with DFOS, we demonstrate the performance of short-term traffic forecasts. The initial condition is set based on the proposed method in section 2.2. Figure 5 shows the $k$–$v$ relation at the MAP model parameters derived from DFOS data, which is used to set the initial condition with the latest observed mean velocity shown in Figure 3 (a) $t = -1$–$0$ minutes. The initial condition for the point sensors case cannot be set using the proposed method. Instead, the final state of the simulation at the MAP parameters of ($V^{BN}$, $p$, $q$, $r$) = (20 km/h, 0.15, 0.30, 0.78) is adopted as the initial condition for forecast simulation in the point sensor case.

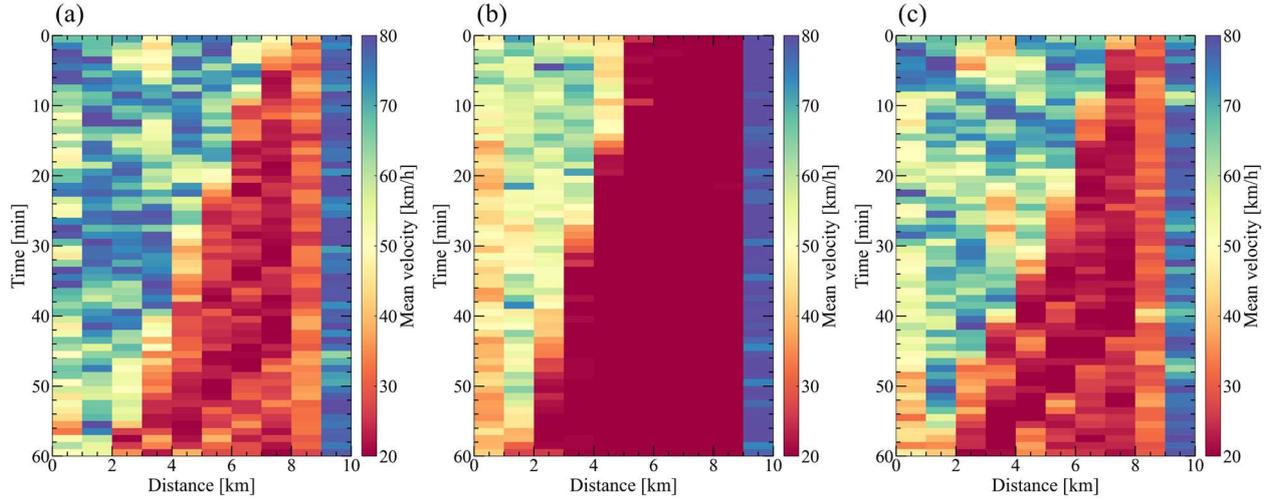

Figure 6. (a) Forecasted mean velocity from DFOS up to 60 minutes. (b) The same as (a) but forecasted from point sensor data. (c) Grand-truth data.

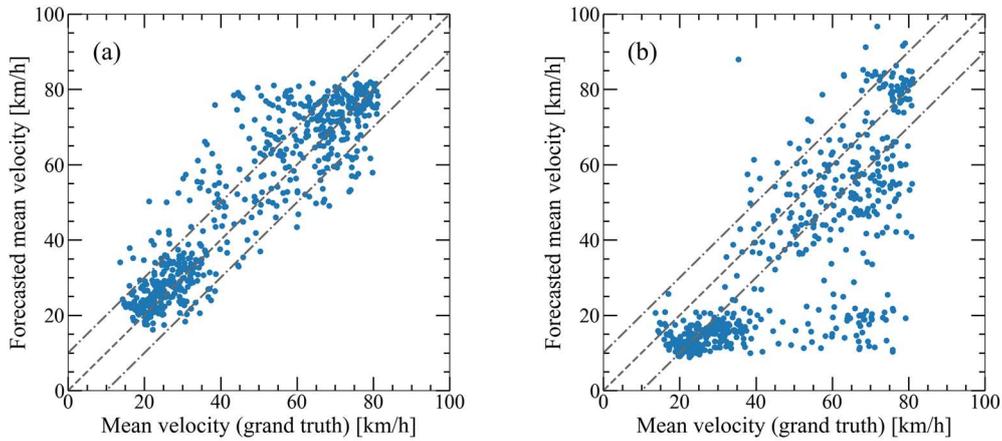

Figure 7. Correlation plots of forecasted and grand-truth mean velocities up to 60 minutes by (a) DFOS and (b) point sensors shown in Figure 6. The orthogonal dashed and dash-dotted lines indicate the 1:1 relation and grand-truth value ±10 km/h, respectively.

TABLE 2.
Representative error indices of forecasted mean velocity.

| Sensor type | Correlation coefficient | MAE | MPE | RMSE |
| --- | --- | --- | --- | --- |
| DFOS | 0.91 | 7.0 km/h | 18% | 9.8 km/h |
| Point sensors | 0.78 | 15 km/h | 35% | 20 km/h |

Figure 6 shows grand-truth future mean velocity and forecasted one from DFOS, point sensors, and grand-truth data. Forecasted mean velocity from DFOS shows a clear correspondence, however, those by point sensors underestimate mean velocities in congestion due to overestimated bottleneck strength (underestimated $V^{BN}$). The anticipation probability $r$ is also underestimated when using the point sensors. Due to this, forecasted mean velocities upstream of growing congestion are underestimated. In addition, the initial condition of the simulation for forecast in the point sensor case is invalid. In the DFOS case, the beginning point of congestion in the 8–9 km section is appropriately reflected in the forecast simulation at $t = 0$–1 minute. In contrast, the forecasted congestion by point sensors at the same time has already grown from 5 to 9 km sections. This is because the final state of the simulation at $t = 0$ minutes at the MAP parameter is invalid. As mentioned above, the bottleneck strength is overestimated when using point sensors. Thus, congestion in the simulation in $t = -20$–0 minutes more grows up than in the actual. Since the final state at $t = 0$ minutes is adopted as the initial condition for the simulation in $t > 0$ minutes, forecasted congestion is overestimated as well as the overestimated bottleneck strength. DFOS can solve these issues because of its visibility along the road.

Correlation plots of grand truth and forecasted mean velocities by DFOS and point sensors are also shown in Figure 7. The correlation coefficient, mean absolute error (MAE), mean percentage error (MPE), and root mean square error (RMSE) of forecasted mean velocities are listed in Table 2. Forecast error of mean velocity by DFOS decreases by approximately half than when using point sensors. The high forecast error of point sensors is caused by two points. One is a misunderstanding of optimal traffic model parameters. The other is an improper initial condition of simulation for forecasting. These issues are due to limited information of point sensors, while DFOS produces continuous information along roads without dead zones.

As prospects of the proposed concepts, validation of actual traffic data is necessary. In addition, synthetic data emulated by sudden lane control due to an accident helps to examine whether the methodology is compatible with such sudden and rare events. As a use case, this real-time traffic forecast system can be applied for adaptive traffic control by road managers to reduce congestion through velocity-controlling vehicles (e.g., [24, 25]), congestion charges, and road pricing (e.g., [26, 27]), and guidance to the optimal route to avoid congestion.

## 5 CONCLUSIONS

We have validated that DFOS enables accurate modeling of traffic flow and a short-term forecast compared to results from conventional point sensors. The proposed methods are verified using pseudo-measured mean velocity data generated by the simulation. The main conclusions are listed below.
1. Our proposed method can estimate the optimal value of the traffic model parameters from their PPDs when using DFOS data. However, optimal values are deviated when point sensors are used. Moreover, the uncertainties of PPDs derived from DFOS are lower than those from point sensors.
2. As an advantage of accurate model parameter estimation, results of the short-term traffic forecasting adopting optimal values of the model parameters are examined. The initial condition is also estimated from the latest observed mean velocities and the $k$–$v$ relation at the optimal model parameters.
3. Forecasted mean velocity by DFOS shows obvious correspondence to the grand-truth data. In contrast, the forecast result by point sensors overestimates the mean velocities of congestion due to overestimated bottleneck strength. The forecast error of mean velocities by DFOS decreases by approximately half compared to the result of point sensors.
4. DFOS overcomes the problem of limited point sensor data and enables accurate data assimilation and traffic forecasting.

## REFERENCES

[1] S. Fleming, "Future of the Environment - Traffic congestion cost the US economy nearly $87 billion in 2018", *World Economic Forum*, available at https://www.weforum.org/agenda/2019/03/traffic-congestion-cost-the-us-economy-nearly-87-billion-in-2018/, May 2019.

[2] K. Lee, M. Eo, E. Jung, Y. Yoon, and W. Rhee, "Short-Term Traffic Prediction With Deep Neural Networks: A Survey", *IEEE Access*, vol. 9, pp. 54739-54756, 2021.

[3] S. Kokubo, J. Tanimoto, and A. Hagishima, "A new Cellular Automata Model including a decelerating damping effect to reproduce Kerner's three-phase theory", *Physica A: Statistical Mechanics and its Applications*, vol. 390, pp. 561-568, 2011.

[4] A. Aw and M. Rascle, "Resurrection of "second order" models of traffic flow", *SIAM Journal on Applied Mathematics*, vol. 60, No. 3, pp. 916–938, 2000.

[5] H. M. Zhang, "A non-equilibrium traffic model devoid of gas-like behavior", *Transportation Research Part B: Methodological*, vol. 36, No. 3, pp. 275–290, 2002.

[6] M. Bando, K. Hasebe, A. Nakayama, A. Shibata, and Y. Sugiyama, "Dynamical model of traffic congestion and numerical simulation", *Physical Review E*, vol. 51, No. 2, pp. 1035-1042, 1995.

[7] K. Nagel and M. Schreckenberg, "A Cellular Automaton Model for Freeway Traffic", *J. Phys. I France*, vol. 2, pp. 2221-2229, 1992.

[8] K. Nishinari, M. Fukui, and A. Schadschneider, "A Stochastic Cellular Automaton Model for Traffic Flow with Multiple Metastable States", *J. Phys. A: Math. Gen.*, vol. 37, pp. 3101–3110, 2004.

[9] H. Fuks and N. Boccara, "Generalized Deterministic Traffic Rules", *Int. J. Mod. Phys. C*, vol. 9, pp.1-12, 1998.

[10] K. Nishinari and D. Takahashi, "Multi-value Cellular Automaton Models and Metastable States in a Congested Phase", *J. Phys. A: Math. Gen.*, vol. 33, pp. 7709-7720, 2000.

[11] M. Takayasu and T. Takayasu, "1/f Noise in a Traffic Model", *Fractals*, vol. 1, pp. 860-866, 1993.
[12] N. Rajewsky, L. Santen, A. Schadschneider, and M. Schreckenberg, "The Asymmetric Exclusion Process: Comparison of Update Procedures", *J. Stat. Phys.*, vol. 92, pp. 151-194, 1998.
[13] S. Sakai, K. Nishinari, and S. Iida, "A New Stochastic Cellular Automaton Model on Traffic Flow and its Jamming Phase Transition", *J. Phys. A: Math. Gen.*, vol. 39, pp. 15327-15339, 2006.
[14] G. A. Wellbrock, T. J. Xia, M.-F. Huang, Y. Chen, M. Salemi, Y.-K. Huang, P. Ji, E. Ip, and T. Wang, "First Field Trial of Sensing Vehicle Speed, Density, and Road Conditions by Using Fiber Carrying High Speed Data", *Optical Fiber Communication Conference Postdeadline Papers*, 2019.
[15] C. Narisetty, T. Hino, M.-F. Huang, R. Ueda, H. Sakurai, A. Tanaka, T. Otani, and T. Ando, "Overcoming Challenges of Distributed Fiber-Optic Sensing for Highway Traffic Monitoring", *Transportation Research Record*, vol. 2675, No. 2, pp. 233-242, 2021.
[16] C. Narisetty, T. Hino, M.-F. Huang, H. Sakurai, T. Ando, and S. Azuma, "TrafficNet: A Deep Neural Network for Traffic Monitoring Using Distributed Fiber-Optic Sensing", *Transportation Research Board 100th Annual Meeting*, 2021.
[17] C. Wiesmeyr, C. Coronel, M. Litzenberger, H. J. Döller, H.-B. Schweiger and G. Calbris, "Distributed Acoustic Sensing for Vehicle Speed and Traffic Flow Estimation", *IEEE International Intelligent Transportation Systems Conference (ITSC)*, pp. 2596-2601, 2021.
[18] H. Prasad, M. Petladwala, D. Ikefuji, H. Sakurai, and M. Otani, "Noisy Trajectory Suppression Method for Traffic Flow Monitoring using Distributed Fiber-Optic Sensing", *The 28th ITS World Congress*, 2022.
[19] H. Prasad, D. Ikefuji, T. Matsushita, M. Takahashi, T. Suzuki, H. Sakurai, and M. Otani, "Localization of Optic Fiber Cables for Traffic Monitoring using DFOS Data", *The 29th ITS World Congress*, 2023.
[20] Z. Ye, W. Wang, X. Wang, F. Yang, F. Peng, K. Yan, H. Kou, and A. Yuan, "Traffic flow and vehicle speed monitoring with the object detection method from the roadside distributed acoustic sensing array", *Frontiers in Earth Science*, vol. 10, 2023.
[21] L. Wang, S. Wang, P. Wang, W. Wang, D. Wang, Y. Wang, and S. Wang, "Traffic Flow and Speed Monitoring Based On Optical Fiber Distributed Acoustic Sensor", 20 Feb. 2024, DOI: https://doi.org/10.48550/arXiv.2402.09422.
[22] R. T. Underwood, "Speed, volume, and density relationship: quality and theory of traffic flow", *Yale Bureau of Highway Traffic*, pp.141-188, 1961.
[23] L. C. Edie, "Discussion of traffic stream measurement and definitions", *Proc. The 2nd International Symposium on the Theory of Traffic Flow*, pp. 139-154, 1963.
[24] A. Kesting, M. Treiber, M. Schönhof, and D. Helbing, "Adaptive cruise control design for active congestion avoidance", *Transportation Research Part C: Emerging Technologies*, vol. 16, pp. 668-683, 2008.
[25] R. Nishi, A. Tomoeda, K. Shimura, and K. Nishinari, "Theory of jam-absorption driving", *Transportation Research Part B: Methodological*, vol. 50, pp. 116-129, 2013.
[26] A. de Palma and R. Lindsey, "Traffic congestion pricing methodologies and technologies", *Transportation Research Part C: Emerging Technologies*, vol. 19, pp. 1377-1399, 2011.
[27] G. Sugiyanto, "The Effect of Congestion Pricing Scheme on the Generalized Cost and Speed of a Motorcycle", *Walailak Journal of Science and Technology*, pp. 95-106, vol. 15, 2016.